\documentclass[floats,floatfix,showpacs,amssymb,prd,twocolumn,superscriptaddress,nofootinbib,nolongbibliography,reprint,preprintnumbers]{revtex4-1}

\usepackage{amssymb,amsmath,verbatim,mathtools,needspace,enumitem,etoolbox,graphicx,physics,microtype,afterpage,xspace,tabularx,lmodern,multirow,bm}
\usepackage{textcomp, gensymb}

\usepackage{braket}

\usepackage{relsize}

\usepackage{dcolumn}

\usepackage[normalem]{ulem}
\usepackage{placeins}
\usepackage{comment}
\usepackage[dvipsnames, usenames]{xcolor}
\usepackage{colortbl}
\definecolor{linkcolor}{rgb}{0.0,0.3,0.5}
\usepackage[unicode, colorlinks=true, linkcolor=linkcolor, citecolor=linkcolor, filecolor=linkcolor, urlcolor=linkcolor, linktocpage, breaklinks]{hyperref}
\usepackage[all]{hypcap}
\usepackage[T1]{fontenc}
\usepackage[utf8]{inputenc}
\usepackage{mathrsfs}
\usepackage[usenames,dvipsnames]{xcolor}
\hypersetup{colorlinks=true,citecolor=romared,linkcolor=romared,urlcolor=romared}

\definecolor{romared}{RGB}{142,0,28}

\newcommand{\be}{\begin{equation}}
\newcommand{\ee}{\end{equation}}

\def\be{\begin{equation}}
\def\ee{\end{equation}}
\newcommand{\beq}{\begin{eqnarray}}
\newcommand{\eeq}{\end{eqnarray}}

\usepackage{aas_macros}
\usepackage{makecell}
\usepackage{soul}
\usepackage[nolist,nohyperlinks]{acronym}
\usepackage{orcidlink}
\usepackage{lipsum}
\usepackage{booktabs}

\acrodef{LSC}[LSC]{LIGO Scientific Collaboration}
\acrodef{BH}{black hole}
\acrodef{NS}{neutron star}
\acrodef{PN}{Post-Newtonian}
\acrodef{BBH}{binary black-hole}
\acrodef{BNS}{binary neutron-star}
\acrodef{NSBH}{neutron-star black-hole}
\acrodef{NR}{numerical relativity}
\acrodef{GW}{gravitational wave}
\acrodef{PSD}{power spectral density}
\acrodef{aLIGO}{Advanced Laser interferometer Gravitational-Wave Observatory}
\acrodef{AZDHP}{aLIGO zero detuned high power density}
\acrodef{GR}{general relativity}
\acrodef{PE}{parameter estimation}
\acrodef{LAL}{LIGO algorithm library}
\acrodef{TPI}{tensor-product interpolant}
\acrodef{SVD}{singular value decomposition}
\acrodef{SNR}{signal-to-noise ratio}
\acrodef{ODE}{ordinary differential equation}
\acrodef{PDE}{partial differential equation}
\acrodef{ROM}{reduced order model}
\acrodef{QNM}{quasi-normal mode}
\acrodef{IMR}{inspiral-merger-ringdown}
\acrodef{LVK}{LIGO-Virgo-KAGRA}
\acrodef{SXS}{Simulating eXtreme Spacetimes}

\newcommand{\ias}{\affiliation{School of Natural Sciences, Institute for Advanced Study, 1 Einstein Drive, Princeton, NJ 08540, USA}}

\newcommand{\orcid}[1]{\href{https://orcid.org/#1}{\includegraphics[width=10pt]{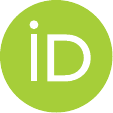}}}
\newcommand{\ben}{\begin{enumerate}}
\newcommand{\een}{\end{enumerate}}

\def\be{\begin{equation}}
\def\ee{\end{equation}}
\def\beq{\begin{eqnarray}}
\def\eeq{\end{eqnarray}}

\def\apjl{\rm{ApJ}}
\def\apjs{\rm{ApJS}}
\def\aap{\rm{A\&A}}

\allowdisplaybreaks

\begin{document}

\pagenumbering{arabic}

\title{How much of the black hole ringdown is sourced within the light ring?}

\author{Mark Ho-Yeuk Cheung \orcid{0000-0002-7767-3428}}
\email{mcheung@ias.edu}
\ias

\date{\today}

\begin{abstract}
Not much.
It has been proposed that applying a rational filter to a black hole ringdown gravitational waveform reveals a direct wave component sourced near the event horizon.
By solving the Teukolsky equation for a point particle plunging into a Kerr black hole, I show that the maximum amplitude of the waves sourced within the prograde light ring is $\lesssim 10\%$ that of those sourced by the full trajectory no matter if the waves are filtered or not.
This agrees with the interpretation of quasinormal modes as waves orbiting the light ring and the direct wave as the power of these modes redistributed in time.
Therefore, both quasinormal modes and direct waves are more a probe of the light ring than the near-horizon region.
For GW250114, I estimate that the component of the waves sourced within the light ring have a signal-to-noise ratio $\rho \lesssim 1$ measured after the peak of the filtered waves.

\end{abstract}

\preprint{000000}

\maketitle

\begin{figure*}
    \centering
    \includegraphics[width=0.5\linewidth]{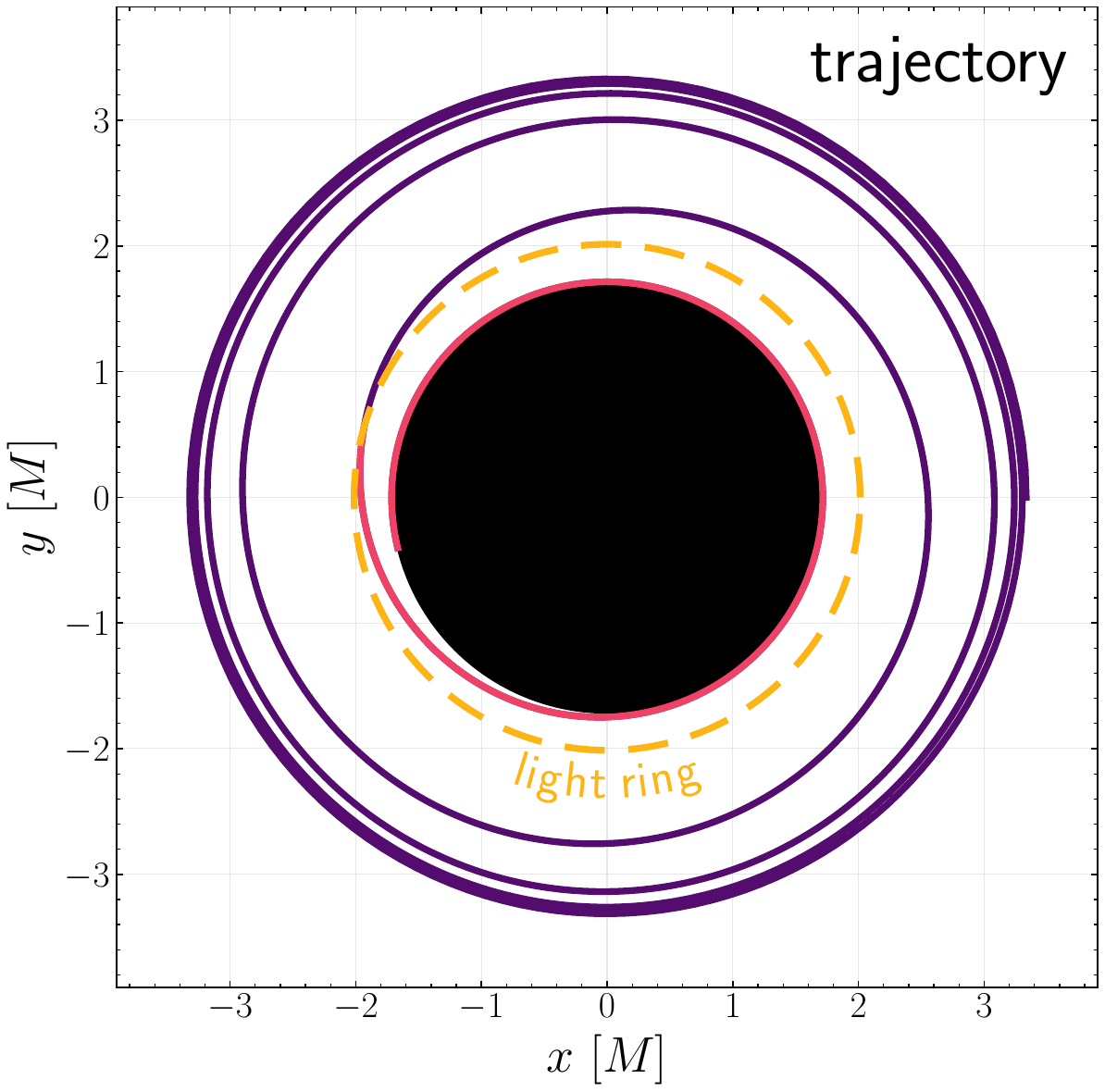} \\
    \includegraphics[width=0.99\linewidth]{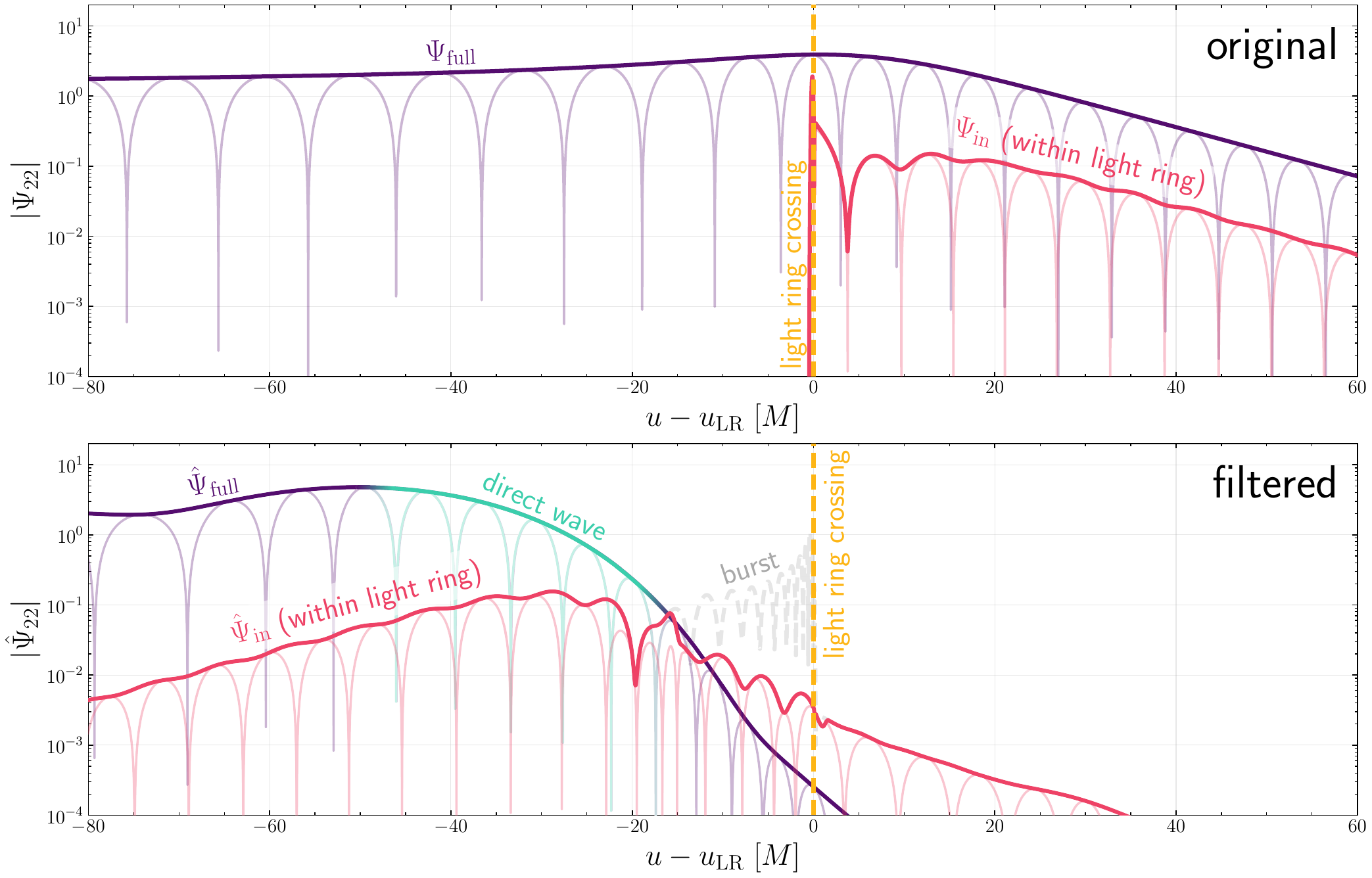}
    \caption{The trajectory (top panel) and first-order perturbation gravitational waveform (bottom panels) of a point particle plunging into a Kerr BH with spin $\chi = 0.7$ from just inside the ISCO following an equatorial geodesic.
    I highlight the part of the filtered waveform that is usually referred to as the direct wave.
    The waves shown are projected onto the ${}_{-2}S_{22}(M\chi\omega_{220}; \theta)$ spin-weighted spheroidal harmonic.
    Due to the abrupt nature of the cut in the trajectory, the waveform sourced within the light ring has an initial burst which causes a spike in the original waveform and a component in the filtered response growing until the light-ring crossing (dashed gray curve in the bottom panel).
    As this component is an artifact introduced by the cut, I remove it with a low-pass filter; see the main text for discussion.
    Disregarding the burst, The maximum amplitude of the waves sourced by the particle when it is within the prograde light ring are $\lesssim 10\%$ that of the full waveform, no matter if filtered or not.
    }
    \label{fig:main}
\end{figure*}

The gravitational waves emitted by a point particle plunging into a Kerr black hole (BH) can be computed with the Teukolsky equation in black hole perturbation theory~\cite{Teukolsky:1973ha, Detweiler:1979xr, Sasaki:1981sx}.
The Green's function of the equation has poles in the (negative-${\rm Im}\,\omega$) lower half plane.
The response after the particle crosses the prograde light ring is dominated by decaying quasinormal modes (QNMs) with negative ${\rm Im}\,\omega$; see Ref.~\cite{Berti:2025hly} for a recent review.
A rational filter~\cite{Ma:2022wpv},
\begin{equation}
    \mathcal{F}(\omega) = \prod_{\ell m n} \frac{\omega - \omega_{\ell m n}}{\omega - \omega^{*}_{\ell m n}} ,
    \label{eq:filter}
\end{equation}
with $\omega_{\ell m n}$ the complex frequencies of QNMs and an asterisk denoting complex conjugation, can be applied to the waveform.
For a Schwarzschild BH, this effectively changes the physics of the problem by flipping the poles of the Green's function to the upper half plane~\cite{Ma:2026hcb}.
The filter is all-pass ($|\mathcal{F}(\omega)| = 1$) and redistributes power within the waveform by transforming the QNMs into an anti-causal component growing exponentially until the onset of the original QNM~\cite{Ma:2022wpv}.
This has been described as effectively removing them from their original time window and revealing the non-QNM content of the waveform buried underneath it. 

When the rational filter is applied to the particle-plunge waveform in a given $\ell m$ harmonic, the few decaying oscillation cycles after the peak of the filtered waveform are referred to as the direct wave~\cite{Oshita:2025qmn,Lu:2025vol}.
For a Schwarzschild BH and $\ell m = 22$, it has been shown to be dominated by the response of the anti-causal poles, which at a given retarded time $u$ can be computed as the sum of integrals of the source term with lower limits at the horizon~\cite{Ma:2026hcb},
\begin{multline}
        \hat{\Psi}^{\rm ac}_{\ell m}(u) = 2\,{\rm Re} \sum_{n} E_{\ell m n} \, e^{-i {\omega}^*_{\ell m n} u} \\ \times \int_{r_+}^{R(u)} \frac{\dd R^\prime}{|\dot{R^\prime}|} \, e^{i {\omega}^*_{\ell m n} t(R^\prime)} \, \mathcal{S}_{\ell m n}(R^\prime) ,
    \label{eq:dwintegral}
\end{multline}
where ``ac'' stands for ``anti-causal'', the hat on top of $\hat{\Psi}^{\rm ac}_{\ell m}$ indicates that it is rational filtered, $E_{\ell m n}$ are the excitation factors, $r_+$ is the Schwarzschild coordinate radius of the (outer) horizon, $t(R^\prime)$ is the Schwarzschild coordinate time when the particle is at $R^\prime$, $R(u)$ is the radial coordinate of the particle at the event whose outgoing radial null ray reaches future null infinity $\mathscr{I}^+$ at retarded time $u$, i.e. the solution of $t(R) - r_*(R) = u$ where $r_*$ is the tortoise coordinate, and $\mathcal{S}_{\ell m n}$ is a function that depends on the point-particle source trajectory.

It was shown for Schwarzschild BHs~\cite{Ma:2026hcb}, and argued heuristically for Kerr BHs~\cite{Oshita:2025qmn,Sun:2026mto}, that when the near-horizon limit $r \to r_+$ is taken for the radial dependence of the source term and the Green's function,
\begin{equation}\label{eq:psi_h}
    \hat{\Psi}^{\rm ac}(u) \sim \hat{\Psi}_{\rm nH} \equiv A_{\rm nH} (u) e^{-i \int^u \omega_{\rm G}(u^\prime) \, \dd u^\prime},
\end{equation}
where ``nH'' is for ``near-horizon'', and
\begin{equation}
    \omega_{\rm G} = \frac{m \left( \Omega + \beta \, \Omega_{\rm H} \right)}{1 + \beta} - i \, \frac{2 \, \kappa \, \beta}{1 + \beta} ,
    \label{eq:omegaG}
\end{equation}
where $\Omega = \dd \Phi / \dd t$ is the orbital frequency of the particle, $\beta = |\dd r_* / \dd t|$ is its coordinate speed in the tortoise coordinate $r_*$, $m$ is the azimuthal number, $\Omega_{\rm H}$ and $\kappa$ are the angular velocity ($= 0$ for Schwarzschild) and the surface gravity of the horizon.
Additionally, if we take the limit $u \to \infty$, we have $\beta \to 1$ and $\Omega \to \Omega_{\rm H}$, so $\omega_{\rm G}$ approaches the horizon mode frequency $\omega_{\rm H}^{(1)} \equiv m \,\Omega_{\rm H} - i \, \kappa$~\cite{Mino:2008at,Oshita:2025qmn}.
In the same limit, $A_{\rm nH} \sim \omega_{\rm G}(u) - \omega^{(1)}_{\rm H} \sim e^{-\kappa u}$~\cite{Oshita:2025qmn,Sun:2026mto}, so the complex frequency of $\hat{\Psi}_{\rm nH}$ approaches $\omega^{(2)}_{\rm H} \equiv m \, \Omega_{\rm H} - 2 \, i \, \kappa$~\cite{Zimmerman:2011dx,Oshita:2025qmn,Sun:2026mto}.
Up to a constant complex amplitude, $\hat{\Psi}_{\rm n H}$ seems to resemble the filtered waveform $\hat{\Psi}$ within the time window where it is usually called the direct wave, i.e. the few cycles after its peak~\cite{Oshita:2025qmn} (from now on I will drop the $\ell m$ subscript and assume $\hat{\Psi} \approx\hat{\Psi}^{\rm ac}$).
Therefore, it has been suggested that direct waves could be a probe of the near-horizon region~\cite{Lu:2025vol}.
Note that the direct wave is defined as a component of the filtered waveform, not of the unfiltered one, and it is not a horizon mode, but merely \textit{approaches} the complex frequency $\omega^{(2)}_H$.
These distinctions seem to have driven active discussion in the field (see, e.g., Refs.~\cite{DeAmicis:2026tus,Kuntz:2026xep,Kankani:2026kst,Ma:2026hcb,Dyer:2026yex,Chung:2026eph,Sun:2026mto}).
Note that the heuristic argument in~\cite{Oshita:2025qmn} for the validity of Eq.~\eqref{eq:psi_h} in the Kerr BH case has been disputed~\cite{Kuntz:2026xep}, but it does not affect the conclusions made here.

Before proceeding, what do we mean by the \textit{near}-horizon region?
Traditionally, both electromagnetic (EM) waves and GWs probe the spacetime around a BH at least as close as the prograde light ring~\cite{Cardoso:2016rao,EventHorizonTelescope:2019dse}.
Therefore, I will use the most lenient definition that preserves a sense of novelty, i.e. that the near-horizon region is defined to be the region \textit{within the prograde light ring}.
If I were to further relax the definition to include regions outside of the light ring, then traditional methods would be classified as near-horizon probes, rendering the term trivial.

We must also define what we mean by a near-horizon \textit{probe}.
While no waves emitted exactly at the horizon will reach us, at least some from the \textit{near}-horizon region will.
With high enough signal-to-noise ratio (SNR), whatever method used to analyze the ringdown will trivially be sensitive to near-horizon physics.
Also, if there are significant qualitative changes to horizon physics, e.g. the horizon is replaced by a reflective barrier~\cite{Price:2017cjr,Cardoso:2017cqb}, most methods of analyzes will be sensitive to these changes. 
Therefore, I adopt a more useful definition, i.e. a near-horizon probe is a method in which, assuming GR, the waves sourced near the horizon can be \textit{readily separated from the rest of the waveform and analyzed in isolation}.
For example, a method is a near-horizon probe if it filters the waveform to give a window of time in which the waveform is predominantly sourced near the horizon, with enough signal content in that window for it to be analyzed in real detector data.
In that case, a deviation of the filtered waveform in that window is likely caused by new near-horizon physics alone.
I will show that the rational filter method cannot be a near-horizon probe in this sense.

To understand how much of the ringdown is sourced within the light ring, I solve the Teukolsky equation~\cite{Teukolsky:1973ha} in the time domain for a particle plunging from the prograde innermost stable circular orbit (ISCO), i.e.\ the Ori--Thorne plunge~\cite{Ori:2000zn}.
I evolve the $s = -2$, $m = 2$ Teukolsky equation on hyperboloidal slices in horizon-penetrating coordinates $\sigma$~\cite{Zenginoglu:2007jw,Racz:2011qu,PanossoMacedo:2023qzp}.
The angular dependence is expanded in spin-weighted spherical harmonics up to $\ell = 11$, and the resulting $\ell$-coupled system is solved in $\sigma$ with a nodal discontinuous Galerkin scheme (following Refs.~\cite{Field:2009kk,Vishal:2023fye}), an exact upwind flux and a fourth-order Runge--Kutta time integrator, with the delta functions in the point-particle source~\cite{Sundararajan:2007jg} imposed as exact jump conditions across the moving position of the particle.
The waveform extracted at future null infinity $\mathscr{I}^+$ is $\Psi \equiv \lim_{r \to \infty} r \, \psi_4$.
To isolate the waveform $\Psi_{\rm in}$ sourced within the light ring, I multiply the whole point-particle source term by a turn-on window $W(R)$ that follows the instantaneous radius $R(t)$ of the particle, which transitions from $0$ to $1$ with width $\delta_r = 0.1 \, (r_{\rm LR} - r_+)$ as the particle crosses the light ring.
Note that I apply this window directly to the time-domain source term.
This is physically different from the procedure described in Eqs. (10) and (11) of Ref.~\cite{Oshita:2025qmn} where the window is applied to the Sasaki--Nakamura (SN) equation~\cite{Sasaki:1981sx} source term, which contains integrals that make it nonlocal in radius~\cite{Watarai:2024huy}.
The software used and documentation of the technical details are publicly available~\cite{cheung_2026_22071387}.

Our goal is to examine the waveform in the $\ell m = 22$ harmonic.
Eq.~\eqref{eq:dwintegral} can only hold approximately for Kerr BHs due to mode mixing, so there is no reason to stay in spherical harmonics.
Therefore, I assume that the waveform is a sum of $_{-2}S_{\ell2}(M\chi \omega_{\ell20};\theta)$ spheroidal harmonic components up to $\ell = 11$ and solve for the contribution of the $_{-2}S_{22}(M\chi \omega_{220};\theta)$ harmonic.
The $\Psi_{\rm full}$ and $\Psi_{\rm in}$ used later are obtained this way.
With such procedure, we eliminate the $\ell20$ QNMs with $\ell > 2$ mixed into the $\ell m = 22$ waveform.

In Fig.~\ref{fig:main} I show the trajectory of the plunging particle and the resulting waveforms for a Kerr BH with spin $\chi = 0.7$.
$u_{\rm LR} $ is defined to be the retarded time at which the particle crosses the prograde light ring.
The waveform $\Psi_{\rm full}$ computed for the full trajectory exhibits the usual inspiral-merger-ringdown behavior.
The waveform $\Psi_{\rm in}$ sourced within the light ring has an initial burst, caused by the rather abrupt cut imposed by the narrow $W(R)$ transition window, followed by the same QNM ringing.
This burst is an artifact canceled by a burst with an opposite amplitude in the waveform sourced outside the light ring $\Psi_{\rm out}$.
While it is possible to remove it by using a gentler transition window in $W(R)$, it would risk removing a significant fraction of the source within the light ring.
Disregarding the sharp burst, $\Psi_{\rm in}$ exhibits QNM ringing at later times and has an amplitude $\lesssim 10\%$ that of the full waveform.
This shows that most of the ringdown QNMs are sourced on or outside the light ring, consistent with the usual interpretation~\cite{Cardoso:2008bp}.

For the rational filter study, we include both prograde and retrograde modes in the $\ell m = 22$ harmonic up to overtone number $n = 7$.
Motivated by Ref.~\cite{Sun:2026mto}, I also repeat all the analyses with a filter that only includes modes up to $n = 2$ and find that none of the conclusions changes.
The rational filter is an all-pass anti-causal filter, so it merely redistributes the power of the initial waveform to earlier times~\cite{Ma:2022wpv,Ma:2026hcb,Sun:2026mto}.
Indeed, as shown in the lower panel of Fig.~\ref{fig:main}, the maximum amplitude of the filtered $\hat{\Psi}_{\rm in}$ (disregarding the burst) remains $\lesssim 10\%$ that of the full waveform.
The filtered response of the initial burst shows up as a growing high-frequency component right before the light-ring crossing time (gray dashed curve in the bottom panel).
In the Fourier domain, while $\tilde{\Psi}_{\rm full}(\omega)$ decays to zero at high frequencies, $\tilde{\Psi}_{\rm in}(\omega)$ and $\tilde{\Psi}_{\rm out}(\omega)$ asymptote to equal and opposite constants.
Therefore, this component is an artifact of the cut, and I remove it with a low-pass filter.
To make sure that I do not affect the part of the waveform usually referred to as the direct wave, I only replace $\hat{\Psi}_{\rm in}$ by the low-passed version between $u - u_{\rm LR} \approx -20 M$ and $0 M$ in the time domain.
The low-passed $\hat{\Psi}_{\rm in}$ is shown as the red curve in the lower panel of Fig.~\ref{fig:main}, and the unadjusted version can be recovered by adding the burst component.
Nonetheless, the conclusions made later do not depend on whether the burst is kept or removed.

The early direct wave part of $\hat{\Psi}_{\rm full}$ has been identified in GW250114~\cite{Lu:2025vol}.
As $\hat{\Psi}_{\rm in}$ is $\lesssim 10\%$ of $\hat{\Psi}_{\rm full}$ there, Ref.~\cite{Lu:2025vol} is more a probe of the light ring than the near-horizon region.
Indeed, converting $\hat{\Psi}$ into the filtered strain $\hat{h}$, rescaling to the amplitude and remnant mass of GW250114 and using the same detector noise~\cite{LIGOScientific:2025rid,LIGOScientific:2025wao}, the network SNR computed starting at the peak of $\hat{h}_{\rm full}$ is $\rho\sim 26$ for $\hat{h}_{\rm full}$ and $\rho\sim 0.7$ for $\hat{h}_{\rm in}$.
These numbers are unchanged regardless of whether I included the burst (gray dashed lines in Fig.~\ref{fig:main}) because it is suppressed in $\hat{h}_{\rm in}$ and high-frequency components contribute negligibly to the SNR.
Even if explicitly modeled, the waves sourced within the light ring would be undetectable with current detector sensitivity.

At later times, i.e. $u - u_{\rm LR}\gtrsim - 20M$, the amplitude of $\hat{\Psi}_{\rm in}$ surpasses that of $\hat{\Psi}_{\rm full}$.
However, this part of the waveform is affected by the filter's response to the burst.
Moreover, it seems to be as late as the ``transition time'' found in Ref.~\cite{Sun:2026mto} when the extraction of direct waves starts to become unreliable.
Therefore, while at later times at higher SNR the direct wave could be a near-horizon probe in principle, it has yet to be shown rigorously.
On the other hand, if we explicitly model the part of the waves sourced within the light ring as a component of the full waveform, then both traditional QNMs and direct waves can test near-horizon physics if the SNR is high enough, but that would not be a ``near-horizon probe'' as defined on the last page.

In the literature, when a filtered waveform is plotted, it is usually shifted forward in time by aligning it with its unfiltered version during the inspiral.
I do not do that here because each $u$ corresponds to a specific location $R(u)$ of the particle on its trajectory, and shifting in time breaks this correspondence.
For Schwarzschild BHs, it was shown that the filtered waveform $\hat{\Psi}(u)$ at $u$ is the integral of the source between the horizon and the particle's instantaneous position $R(u)$~\cite{Ma:2026hcb} (cf. Eq.~\eqref{eq:dwintegral}).
The integration domain is fully within the light ring when $u - u_{\rm LR}>0$, but $\hat{\Psi}_{\rm full}$ has decayed by orders of magnitude already.  
This is consistent with the fact that only a small fraction of the waves is sourced within the light ring.

If direct waves are mostly sourced outside the light ring, why would it resemble $\hat{\Psi}_{\rm nH}$ in Eq.~\eqref{eq:psi_h}~\cite{Oshita:2025qmn}?
The key is that while $r \to r_+$ is taken for the radial dependence of the source term and Green's function, no limit of $u \to \infty$ of the trajectory itself is taken to derive Eq.~\eqref{eq:psi_h}.
The full trajectory of the particle stays intact within the terms $\Omega$ and $\beta$ in $\omega_{\rm G}$, so it is unsurprising that the frequency evolution works at early times.
Indeed, in numerical relativity (NR) simulations, the real frequency of $\hat{\Psi}_{\rm full}$ does not track the $u\to \infty$ limit ${\rm Re}(\omega^{(2)}_{\rm H}) = m \,\Omega_{\rm H} $ well at the times when the direct wave is well resolved, unless the orbital frequency itself is close to that limit in the first place (which is the case at around $\chi = 0.7$)~\cite{Kankani:2026kst,Sun:2026mto}.
On the other hand, what is less trivial is that the complex amplitude modulation of $\hat{\Psi}_{\rm nH}$ seem to track $\hat{\Psi}_{\rm full}$ well over the few cycles identified as the direct wave.
However, this is also unsurprising because both their amplitudes vary on the time scale of the plunge, so their ratio may look approximately constant over a few periods of the wave.
All in all, the fit between $\hat{\Psi}_{\rm full}$ and $\hat{\Psi}_{\rm nH}$ simply means that the direct wave tracks the trajectory of the particle well outside the light ring.

A related claim about direct waves is that they could probe the interior of the ergosphere.
This could be true for $\chi \gtrsim 0.7$, where the ergosphere extends beyond the prograde light ring.
However, in that case the traditional QNMs are already a probe of the ergosphere, so direct waves do not provide straightforward benefits in this regard.

Incidentally, the term ``direct wave'' suggests a component of the waveform that arrives at the observer directly, without orbiting the light ring as QNMs.
Such a name was chosen because that component was thought to have been revealed after the QNMs were removed.
However, a QNM sum is not a valid description of the unfiltered waveform at times as early as $u - u_{\rm LR} \lesssim -20 M$ (see e.g. Refs.~\cite{Ma:2026qbq,Oshita:2026vxh,Su:2026gmp}), so we should not think of the filter as ``removing QNMs and revealing hidden components'' there.
Moreover, the peak of the filtered waveform is just the peak of the redistributed QNM power, which does not necessarily represent non-QNM content that was ``revealed'' under the QNMs.
What is called the direct wave should simply be understood as the few cycles after the peak of the time-redistributed waveform.
This seems to be closer to the interpretation mentioned in the more recent Ref.~\cite{Sun:2026mto}.
For these reasons, we should be careful not to take the terminology of ``direct wave'' literally.
We may even want to stop using such terminology altogether.

The main conclusions hold also for BH spins from $\chi = 0$ to at least $0.95$, and for plunges with other angular momentum values.
Additional figures for these results, convergence tests, and the code and data used here are publicly available~\cite{cheung_2026_22071387}.
Note that the conclusions in this paper might not apply exactly to the case of equal-mass mergers, although I expect the conclusions to carry over at least qualitatively~\cite{Nichols:2011ih}.

I thank Naritaka Oshita for help with reproducing the results in Ref.~\cite{Oshita:2025qmn}.
I am a Croucher Fellow supported by the Croucher Foundation.
I acknowledge support from the Nelson Center for Collaborative Research at the Institute for Advanced Study and the Simons Foundation through Award No. SFI-MPS-BH-00012593-10.
This work used Anvil~\cite{song2022anvil} at Purdue University through allocation PHY250264 from the Advanced Cyberinfrastructure Coordination Ecosystem: Services \& Support (ACCESS) program~\cite{boerner2023access}, which is supported by U.S. National Science Foundation grants No.~2138259, 2138286, 2138307, 2137603, and 2138296.
This work used large language models (LLMs) including \texttt{Claude Fable 5}, \texttt{Claude Opus 4.8} and \texttt{5}, and \texttt{Claude Sonnet 5} to assist with literature review, coding, running scripts, documentation, and refining parts of this text, all closely supervised by me.

\bibliography{ringdown_horizon}

\newpage

\setcounter{equation}{0}
\renewcommand{\theequation}{N\arabic{equation}}
\section*{Appendix: Notes on the rational filter}

The rational filter~\cite{Ma:2022wpv},
\begin{equation}
    \mathcal{F}(\omega) = \prod_{\ell m n} \frac{\omega - \omega_{\ell m n}}{\omega - \omega^{*}_{\ell m n}} ,
    \label{eq:filter_notes}
\end{equation}
(same as Eq.~\eqref{eq:filter} in the main text) is a powerful tool for analyzing the ringdown, but it is somewhat counterintuitive and has caused some confusion in the field.
Here I show four equivalent ways of understanding the rational filter that I find to be helpful.
The first and second ways think of the filter as transforming the physical problem at hand into a different one, which could be unphysical but nevertheless useful.
The third and fourth ways think of the filter as simply transforming the waveform itself.
I find the fourth way to be the most insightful, as it reveals the strength and limits of the rational filter.

\textit{The first way:} this is the way of understanding the filter as explained in~\cite{Ma:2026hcb}.
The original, unfiltered waveform in the frequency domain is, schematically,
\begin{equation}
    \tilde{\Psi}(\omega) \sim \int G(\omega) \mathcal{S}(\omega) dr^\prime,
\end{equation}
where $G(\omega)$ is the Green's function and $\mathcal{S}(\omega)$ is the source term.
As the filter is linear, applying it to the waveform,
\begin{equation}\label{eq:apply_F}
    \tilde{\Psi}(\omega) \to\mathcal{F}(\omega)\tilde{\Psi}(\omega),
\end{equation}
is equivalent to applying it to the Green's function
\begin{equation}
    G(\omega) \to\mathcal{F}(\omega)G(\omega).
\end{equation}
This transforms the physical problem at hand into one where the QNM part of the Green's function is anti-causal, i.e. its poles are flipped from the lower half plane to the upper half plane.
The term ``anti-causal'' means that the wave depends on the source to the \textit{future}, instead of to the past (within the past light cone).
So the filter gives a waveform that corresponds to a new, unphysical manner of propagation, but for the same source.

While the problem is transformed into an unphysical one, it does not mean that the rational filter is not \textit{useful}.
The rational filtered waveform $\hat{\Psi}(t)$ is simply another way for us to write down the full information of the original waveform, which turns out to correspond to the waveform in a universe where perturbations propagate anti-causally with the Green's function $\mathcal{F}(\omega)G(\omega)$ instead of causally with $G(\omega)$.
To complain that $\hat{\Psi}(t)$ is unphysical is like complaining that the Fourier transformed $\tilde{\Psi}(\omega)$ is unphysical simply because it differs from the more intuitive $\Psi(t)$.
Later, when explaining the fourth way of understanding the rational filter, I will elaborate on why it is a useful tool.

\textit{The second way:} similar to the first way, we can think of the filter as being applied to the source term
\begin{equation}
    \mathcal{S}(\omega) \to \mathcal{F}(\omega)\mathcal{S}(\omega).
\end{equation}
This way, the Green's function stays causal and corresponds to the usual way waves are physically propagated in first-order GR, but with a source that is ``smeared out'' by the filter.
The effect on the source in the time domain can be understood with the help of the time domain kernel of $\mathcal{F}(\omega)$ I derive later in Eq.~\eqref{eq:F_tau_2} .
Heuristically, for a point source, this can be thought of as ``smearing out'' the particle ahead of itself along the trajectory, with a modulation $\sim e^{-i \omega^*_{\ell m n} t}$.

\textit{The third way:} this has already been discussed in the literature~\cite{Ma:2026hcb,Sun:2026mto}, i.e. to understand the filter as redistributing the power of the waveform in the time domain.
From Eq.~\eqref{eq:filter_notes}, we have $|\mathcal{F}(\omega)| = 1$ for all $\omega$, so it is an all pass filter that does not remove any power.
Therefore,
\begin{equation}
    \mathcal{F}(\omega) = e^{i \phi(\omega)}.
\end{equation}
In the time domain, a sinusoidal component with frequency $\omega$ would be shifted by the phase delay $\Delta t = \phi(\omega)/\omega$.
Hence, the power of the waveform is redistributed in time in a frequency dependent manner.

\textit{The fourth way:} the final way to understand $\mathcal{F}(\omega)$ is by working in the time domain directly.
This is the way that I find to be the most insightful.
The application of a filter can be understood as a convolution in the time domain, i.e. Eq.~\eqref{eq:apply_F} is equivalent to
\begin{equation}\label{eq:TD_trans}
    \Psi(t) \to \hat{\Psi}(t) \equiv  \dfrac{1}{\sqrt{2 \pi}}\int^{\infty}_{-\infty} F(\tau) \Psi(t - \tau)d\tau,
\end{equation}
where $F(\tau)$ is the time domain kernel of $\mathcal{F}(\omega)$,
\begin{equation}\label{eq:F_tau}
    F(\tau) = \dfrac{1}{\sqrt{2\pi}}\int^{\infty}_{-\infty} \mathcal{F}(\omega)e^{-i \omega \tau} d\omega.
\end{equation}
For simplicity, I will assume that there is only a single mode $\ell m n$ in the filter Eq.~\eqref{eq:filter_notes}, but the results generalize easily.
To perform the integral in Eq.~\eqref{eq:F_tau}, rewrite $\mathcal{F}$ as
\begin{equation}
    \mathcal{F}(\omega) = 1 - \dfrac{2\, i\,{\rm Im}(\omega_{\ell m n})}{\omega - \omega^*_{\ell m n}}.
\end{equation}
Performing the inverse Fourier transform in Eq.~\eqref{eq:F_tau} we get,
\begin{align}
    F(\tau) &= \dfrac{1}{\sqrt{2 \pi}}\left[ \int^{\infty}_{-\infty} e^{-i \omega \tau} d \omega \right. \\*
    &\quad \quad \quad\quad - \left. 2 \, i  \, {\rm Im}(\omega_{\ell m n}) \int^{\infty}_{-\infty} \dfrac{e^{-i \omega \tau}}{ \omega - \omega^*_{\ell m n}} d\omega \right] \nonumber \\
    &= \sqrt{2\pi}\left[ \delta(\tau) + 2 \, {\rm Im} (\omega_{\ell m n}) e^{- i \omega^*_{\ell m n} \tau } \Theta(- \tau)\right], \label{eq:F_tau_2}
\end{align}
where $\Theta$ is the Heaviside function.
Then, plugging Eq.~\eqref{eq:F_tau_2} into Eq.~\eqref{eq:TD_trans}, we have,
\begin{equation}\label{eq:Psi_t}
    \hat{\Psi}(t) = \Psi(t) + 2 \, {\rm Im}(\omega_{\ell m n}) \int^{\infty}_{0} e^{i \omega^*_{\ell m n} \tau} \Psi(t + \tau) d\tau,
\end{equation}
where we changed the dummy variable $\tau \to -\tau$.
Now, note the following useful identity,
\begin{equation}\label{eq:norm}
    \int^{\infty}_{0} e^{i \omega^*_{\ell m n} \tau} e^{- i \omega_{\ell m n}\tau}d\tau = - \dfrac{1}{2 \, {\rm Im}(\omega_{\ell m n})},
\end{equation}
which can be used to replace the $2 \, {\rm Im}(\omega_{\ell m n})$ in front of the integral in Eq.~\eqref{eq:Psi_t}.
Next, define the following,
\begin{align}
    Q_{\ell m n}(\tau) &\equiv e^{- i \omega_{\ell m n} \tau}, \\
    (f \,|\, g)_\tau &\equiv \int^{\infty}_{0} f^*(\tau) g(\tau) d\tau, \label{eq:overlap}\\
    \Psi(t + \tau) &\equiv \Psi(t) \psi(\tau), \quad \text{(for fixed $t$)}. \label{eq:small_psi}
\end{align}
Eq.~\eqref{eq:small_psi} can be understood as follows: for a given fixed time $t$, $\psi(\tau)$ is the original waveform $\Psi$ to the future of $t$ (i.e. starting at $\tau = 0$), but with an amplitude normalized to unity at the start $\tau = 0$.
Then, using all of Eq.~\eqref{eq:Psi_t} to~\eqref{eq:small_psi},
\begin{align}
    \hat{\Psi}(t)  &= \Psi(t) - \dfrac{\left( Q_{\ell m n} | \Psi(t + \tau)\right)_\tau}{\left(Q_{\ell m n} | Q_{\ell m n} \right)_\tau}  \nonumber \\
    &= \Psi(t)\left[ 1 - \dfrac{\left( Q_{\ell m n} | \psi\right)_\tau}{\left(Q_{\ell m n} | Q_{\ell m n} \right)_\tau}\right].\label{eq:filter_TD_final}
\end{align}
The term in the square bracket is a ``mismatch'' term defined by one minus the ``overlap'' norm in Eq.~\eqref{eq:overlap}.
It is now clear what the rational filter does to a waveform $\Psi(t)$ in the time domain: for each time $t$, the filter re-weights the value of $\Psi(t)$ by the mismatch between $\psi(\tau)$ and a unit amplitude QNM $Q_{\ell m n}$, where $\psi(\tau)$ is the waveform to the future of $t$ but rescaled such that it has a unit amplitude at $t$.

Eq.~\eqref{eq:filter_TD_final} shows the power of the rational filter explicitly.
If $\Psi(t) = A_{\ell m n}e^{- i \omega_{\ell m n} t}$, then $\psi(\tau) = Q_{\ell m n}(\tau)$, so the mismatch is zero and $\hat{\Psi}(t) = 0$ for $any$ $A_{\ell m n}$.
If $\Psi(t)$ is instead a QNM with another frequency $\omega_{\ell^\prime m^\prime n^\prime}$, it can be shown from Eq.~\eqref{eq:filter_TD_final} that the mismatch term is constant over time, so the amplitude of the mode gets rescaled~\cite{Ma:2022wpv}.
If $\Psi(t)$ is a power law decay, the mismatch term asymptotes to a pure phase shift as $t \to \infty$.
Therefore, if $\Psi(t)$ is a combination of QNMs and power laws, which is true for intermediate and late times in the ringdown waveform, the filter removes a set of QNMs, rescales the amplitude of other QNMs, and shifts the phase of the power law decay components asymptotically, without requiring the knowledge of the complex amplitude of $any$ of these components.
This is much more accurate and computationally efficient than first performing a least-squares fitting of the QNM component and then removing it from the waveform.
Incidentally, note that for a fixed $t$, $\hat{\Psi}(t)$ does not depend on the original waveform $\Psi(t^\prime)$ for $t^\prime < t$, so the rational filter is indeed anti-causal, i.e. ``forward looking'', so as long as the QNM component is clean for $t > t_0$, it will be removed cleanly in the same range of time no matter what the waveform looks like for $t < t_0$.   

However, if $\Psi(t)$ is not a clean combination of QNMs and power laws, then the effect of the rational filter on $\Psi(t)$ is non-trivial.
For any $\Psi(t)$, we can always define the split
\begin{equation}
    \Psi(t) \equiv \Psi_{\rm clean}(t) + \delta \Psi(t),
\end{equation}
where $\Psi_{\rm clean}(t)$ is a combination of clean QNMs and power law components.
We already know how the filter acts on $\Psi_{\rm clean}(t)$, so by linearity we can simply look at its effect on $\delta \Psi(t)$.
Then, the mismatch term in Eq.~\eqref{eq:filter_TD_final} is nontrivial and time dependent in general, because $\delta \Psi(t)$ is not a combination of QNMs and power laws.
Therefore, in general, we cannot think of the filter as simply ``removing'' QNM components.

For a ringdown waveform, if the source is a delta function pulse, a QNM sum does not converge at early times $t < t_{\rm conv}$ (see, e.g.,~\cite{Ma:2026qbq,Oshita:2026vxh,Su:2026gmp}), so there is a time before which it cannot be described as a clean QNM sum.
For the ringdown phase of a particle plunge, the waveform is a continuous sum of delta functions at different spacetime events along the trajectory, each with a different $t_{\rm conv}$ up to arbitrarily late times, so in principle the waveform is never a clean sum of QNMs and power laws, although it approaches one at late times~\cite{Ma:2026qbq,Su:2026gmp}.
It is instead the sum of a prompt response component and QNM and tail components with amplitudes depending on time.
At times before the peak of $\Psi(t)$, the magnitude of $\delta\Psi(t)$ is comparable to $\Psi_{\rm clean}(t)$~\cite{Ma:2026qbq,Su:2026gmp}, so the effect of the filter is nontrivial, and should simply be interpreted as redistributing power in time instead of removing clean QNM components.

\textit{Summary:} the rational filter has its strengths and its limits. 
The transformation of the waveform into one that corresponds to an unphysical anti-causal Green's function goes against our usual intuition, but it does not mean that the application of a rational filter is inherently problematic. 
If the waveform is a clean combination of QNMs and power law tails, e.g. at intermediate and late times after the peak of the ringdown, then the rational filter is a powerful tool to reveal subdominant components.
For example, the filter was used in Ref.~\cite{Ma:2022wpv} to show convincingly for the first time that quadratic QNMs exist in the ringdown of NR simulations (subsequently verified by Refs.~\cite{Mitman:2022qdl,Cheung:2022rbm}).
However, at earlier times, when the waveform is not a clean combination of QNMs and power laws, the rational filter redistributes power in time nontrivially.
This is why in the main text I argue against interpreting the early decay of the filtered waveform as a direct emission component that is ``revealed'' after ``removing'' the QNMs.

\end{document}